# Intelligent reputation system for safety messages in VANET

**Ghassan Samara**
Department of Computer Science, Faculty of Information Technology, Zarqa University, Zarqa, Jordan

**ABSTRACT**

Nowadays, Vehicle Ad - hoc Nets (VANET) applications have become very important in our lives because VANET provides drivers with safety messages, warnings, and instructions to ensure drivers have a safe and enjoyable journey. VANET Security is one of the hottest topics in computer networks research, Falsifying VANET system information violates VANET safety objectives and may lead to hazardous situations and loss of life. In this paper, an Intelligent Reputation System (IRS) aims to identify attacking vehicles will be proposed; the proposed system will rely on opinion generation, trust value collection, traffic analysis, position based, data collection, and intelligent decision making by utilizing the multi-parameter Greedy Best First algorithm. The results of this research will enhance VANET's safety level and will facilitate the identification of misbehaving vehicles and their messages. The results of the proposed system have also proven to be superior to other reputational systems.

*Corresponding Author:*

Ghassan Samara,
Department of Computer Science,
Faculty of Information Technology,
Zarqa University, Zarqa, Jordan.
Email: gsamara@zu.edu.jo

## 1. INTRODUCTION

Continuous wireless technology developments offer opportunities for the use of these technologies in driving environment improvements to ensure road safety, infotainment and efficient transport. Worldwide deaths are growing dramatically, and a substantial percentage of these deaths are on roads, in 2018, about 1.35 million people were killed worldwide, and over 50 million are injured. In the next few years, these numbers will rise by around 60 percent, unless action is taken [1], as well as other harms such as the loss of time caused by traffic jams. Traffic jam is the worst thing any driver in the world dreams of avoiding, many vehicles that travel can create problems, or even have trouble that must be notified to other vehicles to prevent overcrowded traffic, besides, many vehicles may dispatch inaccurate information or flawed data, and this can even worsen the situation [2-4]. Vehicles receive messages in vehicular ad hoc networks or send many messages, and not every such message needs to be taken into consideration, as not every vehicle has good intent, and some have an Evil attitude.

Vehicle Ad Hoc Networks (VANET) is a wireless network that connects vehicles to each other and allows for Internet access. VANET is a special group of Mobile Ad Hoc Networks (MANETs) in which nodes move freely, so there is no limitation to their mobility. When each node changes its location, it will remain connected, which means that VANETs have highly dynamic topologies. Nodes are communicating with each other in a single hop or multi-hop [5-6].

VANET security should achieve four objectives, ensuring that the information received is correct (authenticity of the information), who claims to be the source (integrity of messages and authentication of the source), unable to identify and track the source of the message (privacy), and the system is robust [7].

Initiatives of recent research supported by governments and automobile producers are seeking to improve transportation system security and efficiency, and "Fake Information" has been one of the main subjects to search. Current studies recommend that the Road Side Unit (RSU), is responsible for the monitoring of vehicle misbehaviour. RSU also manages the certificate, communicate with the Certificate Authority (CA), broadcast warning messages, and communicate with other RSUs. Current technology is subject to a large RSU overhead because the entire Vehicle Network (VN) communication is under RSU's responsibility [8-9].

An intelligent reputational system for identifying attacking vehicles is proposed in this paper. This system will rely on opinion generation, trust value collection, traffic analysis, position-based, data collection, and intelligent decision making by utilizing the multi-parameter Greedy Best First algorithm. This, in turn, improves network performance by ignoring false alarms from misbehaving cars. The VANET system structure which includes vehicles, RSUs, CAs is shown in Figure 1.

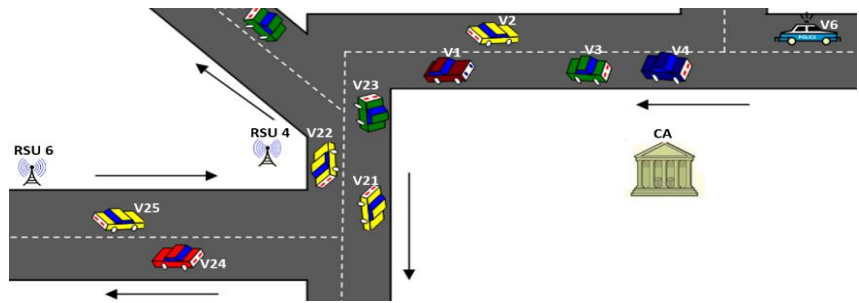

Figure 1. VANET system structure

## 2. LITERATURE REVIEW

In [10] the authors proposed a dynamic reputation system based on events (ERS) which produces confidence according to the behaviour of the vehicle, a Trust Points shall be added to the vehicle every time a specific vehicle sends information matching other information gained from other vehicles, whenever a message matching other messages is received, a single point will raise the value, a vehicle will be treated as trusted when the value hits a predefined threshold, and any message received from it will be processed without inspection, the value will decrease every time it sends incorrect information. While the Vehicle Ad Hoc Reputation System (VARS) proposed by the authors in [11], where the opinion of each vehicle is transmitted to all other vehicles, such information is aggregated to form a reputation for all system vehicles.

In [12-14] The authors discussed and addressed the major safety issues faced by VANET, such as Denial of Service attack, Message Suppression attack, Fabrication Attack, Alteration Attack, Replay Attack, Sybil Attack, Pranksters, Malicious attacker. Furthermore, the authors defined the kinds of VANET attackers such as Selfish Driver. The authors also identified the challenges of VANET safety and security.

In [15], authors proposed a mechanism of identification of the adverse vehicles and offered not to use the Certificate Revocation List (CRL) which is commonly used because it causes delay, overhead processing, and channel interference, the Valid Certificate (VC) or Invalid Certificate (IC), shall apply to each vehicle within the network, and other vehicles on the network shall process messages transmitted from any vehicle depending on these certificates. In [16] the authors proposed a method to deal with car certifications to facilitate the identification of adversarial vehicles. This is done by revoking their certificates, including the Road Side Unit Regulation and the Network Certification Authority. The proposed method aimed to control the network.

In [17] Authors suggested that a reputation server be used as an entity for producing or revoking certificates from untrusted vehicles. The server produces certificates for all vehicles, and it can revoke certificate and stop producing any new when it finds that this vehicle causes problems. This approach is referred to as certified reputation, which is firstly proposed in [18].

In [19] authors proposed a trust-based scheme, the network initially provides a calculation of the trustworthiness of each vehicle. The location data of a vehicle is calculated after trustworthiness has been used. The purpose of calculating confidence is for the greater the value of a node to be trusted, the greater the likelihood of responding with accuracy.

Authors have found a change in the trustworthiness percentage of the data when node numbers are increased, and thus the number of replies is increased. Afterward, authors used the most trusted node's location information. The confidence threshold of any vehicle is above 50%. However, Acceptance of trust data for vehicles with the trustworthiness of more than 50% includes many vehicles and thus high calculations and overheads.

In [20], authors proposed a reputation system-based lightweight message authentication framework and protocol for 5G-enabled vehicular networks (RSMA). Is responsible for managing reputation. A vehicle having a reputation value below the given threshold cannot be approved for participation by the CA; The number of untrusted messages is, therefore decreased from the source in the vehicle networks.

Authors in [21] Authors define the problem of anonymous counting and subsequently propose a categorical distinct pseudo-identity scheme. This scheme accomplishes secrecy to overcome the counting issue. The paper was based on the trust level and vehicle location.

## 3. THE PROPOSED SYSTEM

Basic System assumptions:
1. Each vehicle has an internal memory on board (OBU) that store certificates and information.
2. The Certification Authority (CA) and Road Side Unit (RSU) are two trusted agents.
3. Each vehicle has a unique certificate with a digital signature issued by CA, and cannot be changed over time.
4. All network vehicles sync their clocks with the existing RSU.
5. The decision to accept or reject sender vehicle data must be taken very quickly.

Message and Vehicle types:
Two types of control messages are sent by each car in the system. Periodic beacon (status) messages sent 10 times a second to notify the sender's condition to neighbouring vehicles [12]. Warning messages (event-driven) sent only in situations of danger such as a car crash, ice on the pavement, sudden break. These messages should be sent without alteration or impersonation as quickly as possible [22-23].
System vehicles are categorized into three types [23]: 1- discoverers: Trustful agents with IDs, certificates, vehicle information and those agents in the system are the RSUs and CAs in the existing network. 2- attackers: one or more vehicles performing an attack on other network vehicles. 3- Receivers: Normal vehicles sending and receiving normal data in a network.

Receiving message
Each vehicle has Local Reputation points List (LRL). The list contains an evaluation points which the receiver vehicle calculates for all the vehicles that sent a message to it shortly and RSU Reputation Points List (RRL), shown in Table 1. This is a list calculated by RSU based on the reports sent by network vehicles, and an updated list is distributed to all network vehicles, shown in Table 2.
The three-levels LRL structure are shown in Table 3. Top: that resembles the most reliable vehicles with the highest reputation, this level is at its top so that the vehicle can realize that the sending vehicle is trusted and quickly accept its message. Middle: containing suspicious vehicles that can be malicious and harmful. Bottom: misbehaved (malicious) vehicles that have already sent the recipient false or misleading information.

Table 1. RSU reputation points list structure

| RSU Reputation points List (RRL) |
| --- |
| Top (Low Reputation Points) |
| Middle (Middle Reputation Points) |
| Bottom (high Reputation Points) |

Table 2. LRL update

| ID | Rep. Points | Trust level |
| --- | --- | --- |
| V26 | 13 | Top |
| V18 | 11 | Top |
| V2 | 7 | Medium |
| V57 | 6 | Medium |
| V14 | 4 | Low |
| V11 | 3 | Low |
| V23 | 1 | Low |
| V38 | 1 | Low |

Table 3. Local reputation points list structure

| Local Reputation points List (LRL) |
| --- |
| Top (high Reputation Points) |
| Middle (Middle Reputation Points) |
| Bottom (Low Reputation Points) |

Opinion generation
Each vehicle in the system has its own reputation calculation (Data-centric), and this is done all the time by receiving a number of messages like beacons, warnings or data messages, if a vehicle receives a message of warning from other vehicles, it checks the LRL list first, the good behaviour vehicles on the top of this list, and misbehave vehicles at the bottom. Therefore, if a message came from a vehicle ID situated at the top of a list it will be treated differently from another message received from vehicle located at the bottom. Furthermore, the RRL list plays an important rule and must be taken into consideration in the communication confidence decision since it represents a network opinion on the behavior of the vehicle, while it has fewer weight than LRL since LRL represents a personal experience of the vehicle receiver.

Reputation points:
If the recipient received messages concerning a situation, and one of those messaging contains different information concerning the same situation, the recipient vehicle considers the sender as potential misbehaving and its LRL points are reduced by one, furthermore, the warning about a situation must be sent from vehicle close to that event, and this will mean a false message if the sender is far from the event, when warning of the same situation is received by more than one neighboring vehicle, a message is aggregated, and the LRL for those vehicles increases.

If there are a lot of nearby vehicles, the message can be a false alarm if the warning is only received from one vehicle, the receiver vehicle will then search the sender ID in their LRLs to decide whether or not to accept this information, as shown in Figure 2 for example: In Figure 2 example, each vehicle has ID and heuristic (H). The network consists of nine vehicles, vehicle id = 03 in the middle is the receiver which receives messages from all the neighbors, and its LRL will be as follows:

The network in Figure 2 is consist of 9 vehicles. Every vehicle with ID and heuristic (H). Vehicle Id = 03 in the middle is the receiver receiving messages from the entire neighbourhood, and its LRL is as follows:

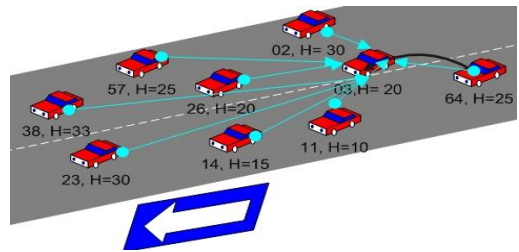

Figure 2. Reputation system example

Equation (1) calculates the threshold

$$Th_{(length)} = \frac{Max\ point - Min\ point}{3} \tag{1}$$

Where $Th_{length}$ is the threshold length, Max point is the maximum reputation point, Min point is the minimum reputation point. The threshold for the previous example will be $Th = \frac{13-1}{3} = 4$. This means that the lowest trust level will be

Trust levels = $\min point - 1 + th$

Then 1-1+4, so the first trust level (low) will be one point to four reputation points, the second trust level (medium), from five reputation points to nine reputation points, the third trust level (top), from ten reputation points to thirteen reputation points, as shown in Table 4.

Table 4. Reputation points

| Trust Level | Threshold |
|---|---|
| Top | 10 - 13 |
| Medium | 5 - 9 |
| Low | 1 - 4 |

is the Strait-Line Distance heuristic between the transmitter vehicle and the event and can be obtained at any time, where each ten meters resembles 1H. So, vehicle ID= 11 is having H=11 as it reported about the event and it is away from it 110 m as SLD, a sending signal strength can calculate the location of the vehicle so that vehicles located far from the event receive greater heuristics and vehicles located near to the event have fewer and more confidence.

H is classified according to (2) for the neighbouring vehicle:

$$H_{Evalation} = \left(\frac{Max_H - Min_H}{3}\right) * 2 \tag{2}$$

Where $Max_H$ is the maximum heuristic for all the neighbors, $Min_H$ is the minimum heuristic for all the neighbors, divided by three to give three levels (Near, Middle, Away) from the event. So the HEvaluation of the sender is calculated when the recipient receives a message. For the previous figure example, the $H_{Evaluation}$ will be $H_{Evaluation} = \left(\frac{33-10}{3}\right) * 2$ will equal 15.34, so vehicles with fewer than 16 are near the event. Vehicles between 16 and 24 are mid - heuristic, the rest high heuristic, more accurate data is usually provided for having better decision nodes with lower heuristics. So LRL after adding the $H_{Evaluation}$ will be, as shown in Table 5 and 6.

Table 5. LRL with Heuristic

| ID  | Rep. Points | Trust level | Heuristic |
|-----|-------------|-------------|-----------|
| V26 | 13          | Top         | Middle    |
| V11 | 11          | Top         | Near      |
| V14 | 7           | Medium      | Near      |
| V57 | 6           | Medium      | Away      |
| V64 | 4           | Low         | Away      |
| V2  | 3           | Low         | Away      |
| V23 | 1           | Low         | Away      |
| V38 | 1           | Low         | Away      |

Table 6. LRL and RRL decision

| LRL    | RRL    | Trust Decision      |
|--------|--------|---------------------|
| Top    | Bottom | High trust (accept) |
| Top    | Middle | High trust (accept) |
| Top    | Top    | Low trust (reject)  |
| Middle | Bottom | High trust (accept) |
| Middle | Middle | Unsure              |
| Middle | Top    | Low trust (reject)  |
| Bottom | Bottom | Low trust (reject)  |
| Bottom | Middle | Low trust (reject)  |
| Bottom | Top    | Unsure              |

So When the vehicle receives a message from just one vehicle about a situation, the recipient is going to verify its LRL and the message is accepted if the sender is at the Top - trusted level. If the sender is at the level of medium or low trust, the vehicle will inspect the most recent RRL received from RSU. If the sender is located to the Top trusted level and the HEvaluation is Near or at least Middle the message will be accepted, But if the sender resides in the medium or low level of confidence, this means the vehicle had previously claimed false information in the network. It sends information alone at the moment, so this information is certainly false. This message is usually maintained for a while and later ignored if the receiver does not receive the same warning from a different vehicle in the network later on, and the sender's reputation points will be reduced, receiver vehicle will send a warning message encrypted by the PKI of the RSU and digitally signed, to the Road Side Unit (RSU), which is a small stations router on streets, to report that this vehicle is still distributing false information in the network from the example above, the most trusted vehicle is V11. Shown in algorithm 1 for a trust decision.

Algorithm 1. Receive a message and calculate trust points

To – Calculate trust points
[ message with data ]
If the same message is received by more than one vehicle:
{ Message is safe, increse rep. points}
Else if the message is receive by only one vehicle:
{ Calculate threshold (from LRL)
let Thre.= ( Max rep. points – Min rep. points )/ 3.
Top = Max rep. points – Thre. { Max rep. points to Top}
Middle = Top – Thre. {Top to Middle}
Low = { Middle to Low)
Compute $H_{Evaluation}$ =((MaxH -MinH)/3)*2.
Categurizt the heuristic.
If sender rep. points within Top and $H_{Evaluation}$ is Near then
{Accept}
Else if sender rep. points within Middle or Low then
{ See RRL, If rep. point within Top or Middle:
{ Ignore
Report to RSU (algorithm 2)
Decrease rep. points}}
Else { Accept }}

Upon receiving a misbehave vehicle warning, RSU will check if the sender is an already a misbehave vehicle or not, if the id of the sender were verified as potential misbehave vehicle, the message would be ignored without processing, once a misbehave vehicle warning is received, RSU checks if the sender is already a misbehave vehicle or not, if the sender's identification has been verified as a potential misbehave vehicle, the message would not be processed, else, two possibilities to process the message. First, only one vehicle sends that specific warning, and then the sender Id and the vehicle Id referred to in the message are regarded as suspect. If a further warning of any of those two vehicles were later reported, the vehicle would formally be regarded as misbehaving vehicle, thus increasing its misbehave points, this vehicle is known to RSU as a misbehave vehicle, the points of misbehavior increases by one point. This vehicle can advance to the top of the lists. Second, if the warning about is a particular vehicle, then the misbehavior points for the vehicle reported shall be increased. Shown in algorithm 2 for a trust decision and shown in Figure 3 for message receive processing.

Algorithm 2. Trust decision

To – RSU update information
[ message with alarm ]
If sender rep. points within Top then
{Ignore}
Else if the warning is generated by only one vehicle:
{ Sender and vehicle reported are suspecious until same aklarm is received by
another vehicle :
{
Both senders are trusted, increase rep. points
reported vehicle is misbehave, Decrease rep. points}}

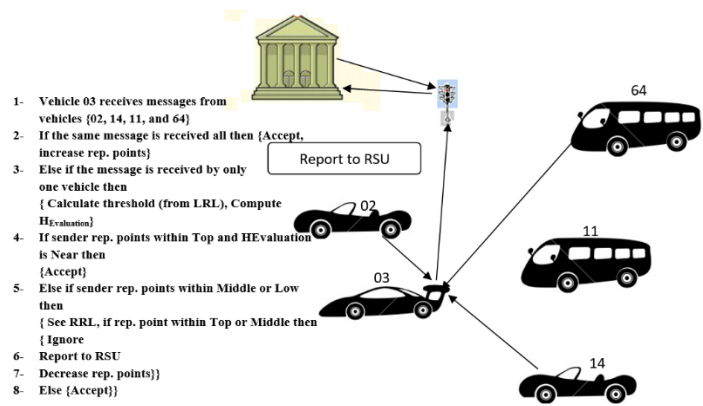

Figure 3. Message receive processing

In case the vehicle enters a new network with the local list LRL not containing current network cars, the vehicle receives RRL from the RSU and accepts their data, and follows RRL data when new suspicious messaging is received from neighbors. In the meantime it generates and builds its own LRL list. LRL should be regularly updated, especially if the network is highly mobile and the many vehicles leave or enter the network within a short period, then the vehicle of the network is rapidly changeable and the RRL according to equation two should, therefore, be updated more frequently.

$$\text{Update} - \text{Thr}\ .= \sum RRL_{vehicles} < \frac{\sum Niehbours}{2} \qquad (3)$$

RSU frequently broadcasts the IDs of the misbehave vehicles for the current network. RSU should send trusted data encrypted with its key. Information is forwarded to the next RSU in the direction the vehicle is moving to, And RSU recipient will update and broadcast the misbehavior information to its neighbors, i.e., RSU and vehicles, on a regular basis.

## 4. SIMULATION AND RESULTS

To prove the accuracy of the proposed protocol, an intensive simulation was conducted with the latest MATLAB R2018a version [24]. The MATLAB R2018a includes the new and better interactive wireless communication environment, the same environment was created with the same parameters, implementation focused on demonstrating the improvement in performance of the proposed system compared to CDPD [21] and RSMA [20]. The simulation parameters are shown in Table 7 for the whole experiment, some parameters are taken from [25].

The CDPD has investigated the effect of distance on the number of trusted messages received. In Figure 4, the results show that when the distance is short, the number of trusted messages is almost 65%. This number decreases as the distance grows, the number of trusted messages, when the distance reaches 160 meters, is very low and scores 25%, while the heuristic helps in avoiding untrusted messages in the proposed IRS protocol and therefore achieve better results, 95% of the messages are trusted on close distances and more than 50% of the messages received on 160 meters, this also means that the efficiency is high, the processing is more focused on trusted messages, and the performance waste is extremely low.

Table 7. Simulation parameters

| Parameter | Value |
|---|---|
| Simulation Grid | 1000 x 1000 |
| Simulation time | 300 sec |
| Vehicle speed | 15–45m/s |
| Number of vehicles Maximum | 100 |
| Number of lanes | 6 (3 in each direction) |
| Scenario | Two-way highway |
| Network interface | Phy/WirelessPhyExt |
| MAC interface | Mac/802 11Ext |
| Interface queue | Queue/DSRC |
| Propagation model | Propagation/Nakagami |
| Number of TDMA slots/frames | 10 |
| Time slot | 2.5ms |
| Message size (safety) | 100 bytes |
| Message size (nonesafety) | 512bytes |
| Transmission range | 300 m, 500m |
| Modulation type | BPSK |
| Antenna type | Antenna/omniantenna |
| Channel type | Channel/wireless channel |
| Data transfer rate | 6, 12, 18, 27Mbps |
| Minimum beaconing interval | 100ms |
| Maximum beaconing interval | 500ms |

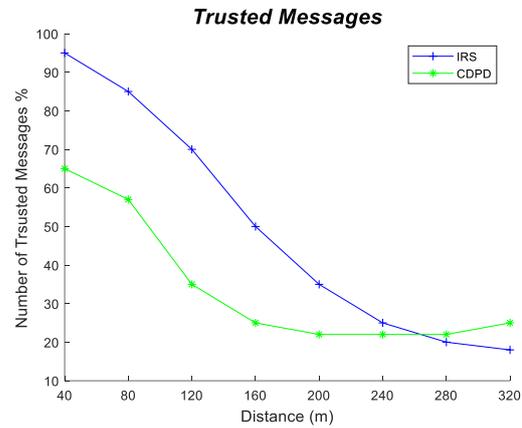

Figure 4. Trusted messages

In Figure 5, The experiment shows the number of vehicles receiving wrong information and believes it to be true. Implementation compared the CDPD with IRS performance, the number of experimental vehicles is 100, CDPD begins suffering from malicious nodes when 5 out of 100 malicious vehicles are present, which means that 5% of vehicles affect another ten vehicles from the remaining network cars. (10/95= 11%), when the number of malicious vehicles is 10 (10%), the number of fooled vehicles reaches 15, which means that 17% of the system vehicles received false messages and believed these messages to be true, while the system implementing IRS starts suffering from malicious nodes when there are only eight attackers and fools three vehicles (3%). When the attackers are ten vehicles, the number of victims reaches 5 (6%).

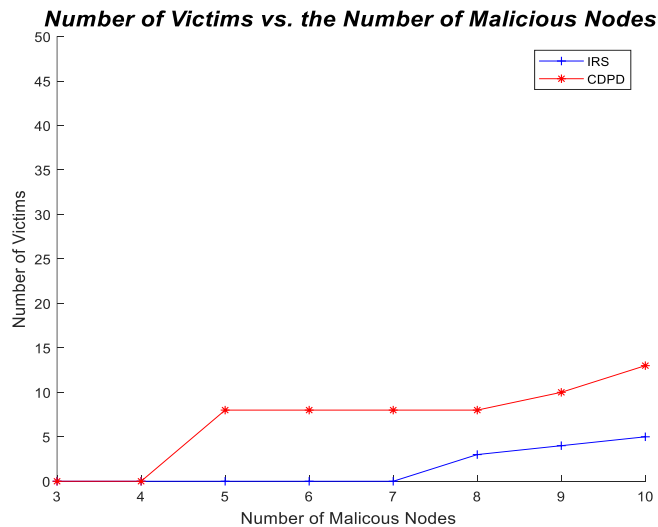

Figure 5. Number of victims vs. the number of malicious nodes.

In Figure 6, the performance of the proposed IRS Protocol against RSMA [20] is examined, where the test concentrated on the execution time under heavy message transaction. The experiment includes the message sent, received, sender verification and evaluation, message categorization, and decision (Accept or Reject). The results showed that, when compared to the RSMA protocol, the proposed system delivers a reasonable performance in terms of execution time.

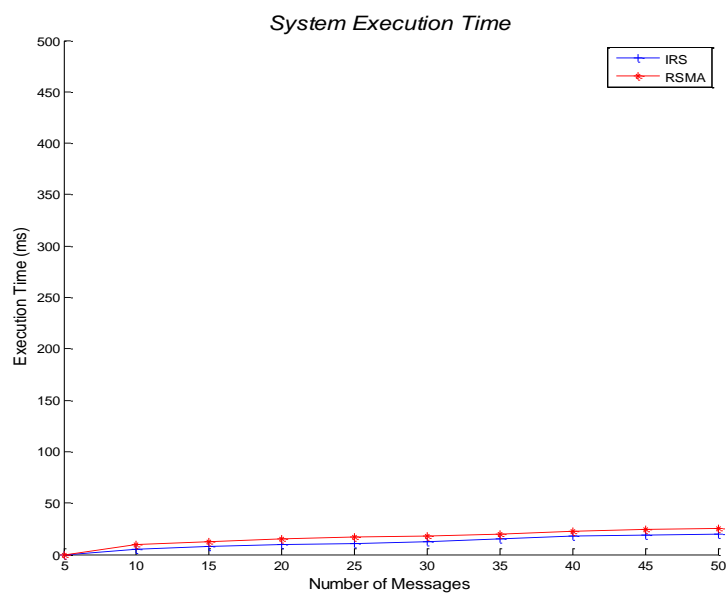

Figure 6. System execution time

## 5. CONCLUSION

Intelligent Reputation System (IRS) protocol was proposed and tested in this paper. The VANET protocol intended to improve security level and to simplify the identification of misbehaving vehicles and their messages. The proposed system relied on opinion generation, trust value collection, traffic analysis, position-based, data collection, and intelligent decision making by utilizing the multi-parameter Greedy Best First algorithm.

The results obtained from this research compared with two protocols CDBD and RSMA by deploying an extensive experiment on MATLAB, the results for the IRS proved a robust and smooth system in terms of the message received, execution time and the number of victims affected.

## ACKNOWLEDGEMENTS

"This research is funded by the Deanship of Research in Zarqa University /Jordan."

## REFERENCES

[1] "WHO | Global status report on road safety 2018," WHO, 2019.
[2] Pattanayak, B.K., Pattnaik, O. and Pani, S., A Novel Approach to Detection of and Protection from Sybil Attack in VANET. In Advances in Intelligent Computing and Communication (pp. 240-247). *Springer*. 2020.
[3] Shrestha, R., Bajracharya, R., Shrestha, A.P. and Nam, S.Y., A new type of blockchain for secure message exchange in VANET. Digital Communications and Networks. 2019.
[4] Liu, X., Huang, H., Xiao, F. and Ma, Z., A blockchain-based trust management with conditional privacy-preserving announcement scheme for VANETs. *IEEE Internet of Things Journal*. 2019.
[5] Samara, G., Alsalihy, W.A.H.A. and Ramadass, S., Increase Emergency Message Reception in VANET. *Journal of Applied Sciences*, 11(14), pp.2606-2612, 2011.
[6] Samara, G., Alsalihy, W.A.A. and Ramadass, S., Increasing Network Visibility Using Coded Repetition Beacon Piggybacking. *World Applied Sciences Journal*, 13(1), pp.100-108, 2011.
[7] Wahid, A., Yasmeen, H., Shah, M.A., Alam, M. and Shah, S.C., Holistic approach for coupling privacy with safety in VANETs. *Computer Networks*, 148, pp.214-230, 2019.

[8] Moreira, E., An evaluation of reputation with regard to the opportunistic forwarding of messages in VANETs. *EURASIP Journal on Wireless Communications and Networking*, (1), p.204, 2019.
[9] Bermad, N., Zemmoudj, S. and Omar, M., Context-aware negotiation, reputation and priority traffic light management protocols for VANET-based smart cities. *Telecommunication Systems*, 72(1), pp.131-153, 2019.
[10] N.-W. Lo and H.-C. Tsai, "A Reputation System for Traffic Safety Event on Vehicular Ad Hoc Networks," *EURASIP J. Wirel. Commun. Netw*., vol. 2009, no. 1, p. 125348, 2009.
[11] F. Dotzer, L. Fischer, and P. Magiera, "VARS: A Vehicle Ad-Hoc Network Reputation System," in *Sixth IEEE International Symposium on a World of Wireless Mobile and Multimedia Networks*, pp. 454-456.
[12] G. Samara, W. A. H. Al-Salihy, and R. Sures, *"Security analysis of Vehicular Ad Hoc Networks (VANET),"* in Proceedings - 2nd International Conference on Network Applications, Protocols and Services, NETAPPS 2010, 2010.
[13] G. Samara, W. A. H. Al-Salihy, and R. Sures, *"Security issues and challenges of vehicular ad hoc networks (VANET),"* in NISS2010 - 4th International Conference on New Trends in Information Science and Service Science, 2010.
[14] Arif, M., Wang, G., Bhuiyan, M.Z.A., Wang, T. and Chen, J., A survey on security attacks in VANETs: Communication, applications and challenges. *Vehicular Communications*, p.100179, 2019.
[15] G. Samara and W. A. H. A. Alsalihy, *"A new security mechanism for vehicular communication networks,"* in Proceedings 2012 International Conference on Cyber Security, Cyber Warfare and Digital Forensic, CyberSec 2012.
[16] G. Samara, W. A. H. Al-Salihy, and R. Sures, *"Efficient certificate management in VANET,"* in Proceedings of the 2010 2nd International Conference on Future Computer and Communication, ICFCC 2010, vol. 3, 2010.
[17] Qin Li, A. Malip, K. M. Martin, Siaw-Lynn Ng, and Jie Zhang, "A Reputation-Based Announcement Scheme for VANETs," *IEEE Trans. Veh. Technol*., vol. 61, no. 9, pp. 4095-4108, Nov. 2012.
[18] T. D. Huynh, N. R. Jennings, and N. R. Shadbolt, *"Certified Reputation - How an Agent Can Trust a Stranger,"* Fifth Int. Jt. Conf. Auton. Agents Multiagent Syst. (AAMAS 2006), pp. 1217-1224, 2006.
[19] S. Das, I. Das, R. P. Singh, P. Johri, and A. Kumar, "Trust-Based Scheme for Location Finding in VANETs Using Trustworthiness of Node," *Springer*, Singapore, pp. 43-55, 2019.
[20] J. Cui, X. Zhang, H. Zhong, Z. Ying, and L. Liu, "RSMA: Reputation System-based Lightweight Message Authentication Framework and Protocol for 5G-enabled Vehicular Networks," *IEEE Internet Things J*., pp. 1-1, 2019.
[21] C. Y. Yeung, L. C. K. Hui, T. W. Chim, S.-M. Yiu, G. Zeng, and J. Chen, "Anonymous Counting Problem in Trust Level Warning System for VANET," *IEEE Trans. Veh. Technol*., vol. 68, no. 1, pp. 34–48, Jan. 2019.
[22] G. Samara, "An intelligent routing protocol in VANET," *Int. J. Ad Hoc Ubiquitous Comput*., vol. 29, no. 1/2, p. 77, 2018.
[23] G. Primiero, A. Martorana, and J. Tagliabue, "Simulation of a Trust and Reputation Based Mitigation Protocol for a Black Hole Style Attack on VANETs," in *2018 IEEE European Symposium on Security and Privacy Workshops (EuroS&PW)*, pp. 127-135, 2018.
[24] "Wireless Communications - MATLAB & Simulink Solutions - MATLAB & Simulink." [Online]. Available: https://uk.mathworks.com/solutions/wireless-communications.html. [Accessed: 11-Apr-2017].
[25] Samara, G., An improved CF-MAC protocol for VANET. *International Journal of Electrical & Computer Engineering (2088-8708)*, 9, 2019.

## BIOGRAPHY OF AUTHOR

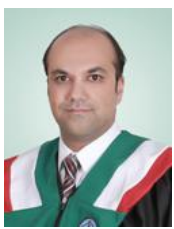

Ghassan Samara: Holds BSc. and MSc. in Computer Science, and PhD in Computer Networks. He obtained his PhD, from Universiti Sains Malaysia (USM) in 2012. His field of specialization is Cryptography, Authentication, Computer Networks, Computer Data and Network Security, Wireless Networks, Vehicular Networks, Inter-vehicle Networks, Car to Car Communication, Certificates, Certificate Revocation, QoS, Emergency Safety Systems. Currently, Dr. Samara is the assistant professor at Zarqa University, Jordan.